\documentclass[oneside,reqno,12pt]{amsart}

\newcommand{\GB}{\mathrm{GB}}
\newcommand{\SGB}{\mathrm{SGB}}
\newcommand{\udu}[4]{{\smash{{{#1}^{#2}}_{#3}}}^{#4}}

\begin{document}

\vspace*{-0.57in}
\begin{flushleft}
\scriptsize{HEP-TH/9509054, UPR-680T}
\end{flushleft}

\vspace*{0.47in}

\title[SUSY Breaking in Four-Dimensional Supergravity]{Higher-Derivative
Gravitation and a New Mechanism \\ for Supersymmetry Breaking in
Four-Dimensions}
\author{Ahmed Hindawi, Burt A. Ovrut, and Daniel Waldram}
\thanks{Talk presented at the Yukawa International Seminar '95, Yukawa
Institute for Theoretical Physics, Kyoto University, Kyoto, Japan,
August 21-25, 1995. Work supported in part by DOE Contract DOE-AC02-76-
ERO-3071 and NATO Grant CRG. 940784.}
\maketitle
\vspace*{-0.3in}
\begin{center}
\small{\textit{Department of Physics, University of Pennsylvania}} \\
\small{\textit{Philadelphia, PA 19104-6396, USA}}
\end{center}
\begin{abstract}

A discussion of the number of degrees of freedom, and their dynamical
properties, in higher derivative gravitational theories is presented.
The complete non-linear sigma model for these degrees of freedom is
exhibited using the method of auxiliary fields. As a by-product we
present a consistent non-linear coupling of a spin-2 tensor to
gravitation. It is shown that non-vanishing
$(C_{\mu\nu\alpha\beta})^{2}$ terms arise in $N=1$, $D=4$ superstring
Lagrangians due to one-loop radiative corrections with light-field
internal lines. We discuss the general form of quadratic $(1,1)$
supergravity in two dimensions, and show that this theory is equivalent
to two scalar supermultiplets coupled to the usual Einstein
supergravity. It is demonstrated that the theory possesses stable vacua
with vanishing cosmological constant which spontaneously break
supersymmetry. We then generalize this result to $N=1$ supergravity in
four dimensions. Specifically, we demonstrate that a class of higher
derivative supergravity theories is equivalent to two chiral
supermultiplets coupled in a specific way to Einstein supergravity.
These theories are shown to possess stable vacuum states with vanishing
cosmological constant which spontaneously break the $N=1$ supersymmetry.

\end{abstract}

\thispagestyle{empty}

\renewcommand{\baselinestretch}{1.2} \large{} \normalsize{}

\vspace*{\baselineskip}

\section{Bosonic Gravitation}

The usual Einstein theory of gravitation involves a symmetric tensor
$g_{\mu\nu}$, the dynamics of which is determined by the Lagrangian
\begin{equation}
\mathcal{L}=\frac{1}{2\kappa^2}\mathcal{R}.
\end{equation}
The diffeomorphic gauge group reduces the number of degrees of freedom
from ten down to six. Einstein's equations further reduce the degrees of
freedom to two, which correspond to a physical spin-2 massless graviton.
Now let us consider an extension of Einstein's theory by including terms
in the action which are quadratic in the curvature tensors. This
extended Lagrangian is given by
\begin{equation}
\mathcal{L} = \frac{1}{2\kappa^2}\mathcal{R} + \alpha\mathcal{R}^2 +
\beta(C_{\mu\nu\alpha\beta})^2 + \gamma(\mathcal{R}_{\mu\nu})^2,
\end{equation}
where $\mathcal{R}^2$, $(C_{\mu\nu\alpha\beta})^2$, and
$(\mathcal{R}_{\mu\nu})^2$ are a complete set of CP-even quadratic
curvature terms. The topological Gauss-Bonnet term is given by
\begin{equation}
\label{GB}
\GB=(C_{\mu\nu\alpha\beta})^2 - 2(\mathcal{R}_{\mu\nu})^2 +
\frac{2}{3}\mathcal{R}^2.
\end{equation}
Therefore, we can write
\begin{equation}
\mathcal{L} = \frac{1}{2\kappa^2} \mathcal{R} + a \mathcal{R}^2 - b
(C_{\mu\nu\alpha\beta})^2 + c\, \GB.
\end{equation}
In this case, it can be shown \cite{A} that there is still a physical
spin-2 massless graviton in the spectrum. However, the addition of the
$\mathcal{R}^2$ term introduces a new physical spin-0 scalar, $\phi$,
with mass $m=(12a\kappa^2)^{-1/2}$. Similarly, the
$(C_{\mu\nu\alpha\beta})^2$ term introduces a spin-2 symmetric tensor,
$\phi_{\mu\nu}$, with mass $m=(4b\kappa^2)^{-1/2}$ but this field,
having wrong sign kinetic energy, is ghost-like. The $\GB$ term, being
purely topological, is a total divergence and does not lead to any new
degrees of freedom. The scalar $\phi$ is perfectly physical and can lead
to very interesting new physics \cite{B}. The new tensor
$\phi_{\mu\nu}$, however, appears to be problematical. There have been a
number of attempts to show that the ghost-like behavior of
$\phi_{\mu\nu}$ is illusory, being an artifact of linearization
\cite{C}. Other authors have pointed out that since the mass of
$\phi_{\mu\nu}$ is near the Planck scale, other Planck-scale physics may
come in to correct the situation \cite{D}. In all these attempts, the
gravitational theories being discussed were not necessarily consistent
and well defined. However, in recent years, superstring theories have
emerged as finite, unitary theories of gravitation. Superstrings,
therefore, are an ideal laboratory for exploring the issue of the
ghost-like behavior of $\phi_{\mu\nu}$, as well as for asking whether
the scalar $\phi$ occurs in the superstring Lagrangian. Hence, we want
to explore the question ``\textit{Do quadratic gravitation terms appear
in the $N=1$, $D=4$ superstring Lagrangian?}''

Before doing this, however, we would like to present further details of
the emergence of the new degrees of freedom in quadratic gravitation. We
begin by adding to Einstein gravitation, quadratic terms associated with
the scalar curvature only. That is, we consider the action
\begin{equation}
\label{R2}
\mathcal{S} = \int d^4x\sqrt{-g}\left(\mathcal{R} + \frac{1}{6} m^{-2}
\mathcal{R}^2 \right).
\end{equation}
The equations of motion derived from this action are of fourth order and
their physical meaning is somewhat obscure. These equations can be
reduced to second order, and their physical content illuminated, by
introducing an auxiliary field $\phi$. The action then becomes
\begin{align}
\label{R2phi}
\mathcal{S} &= \int{d^4x \sqrt{-g} \left(\mathcal{R} +
\frac{1}{6} m^{-2} \mathcal{R}^2 - \frac{1}{6} m^{-2} \left[\mathcal{R}
- 3m^2 \left\{e^\phi - 1 \right\}\right]^2\right)} \\
\label{Rphi}
&= \int{d^{4}x\sqrt{-g}\left(e^{\phi}\mathcal{R} -
\frac{3}{2}m^{2}\left[e^{\phi}-1 \right]^{2}\right)}.
\end{align}
Note that the $\phi$ equation of motion sets the square bracket in
equation \eqref{R2phi} to zero. Hence, action \eqref{Rphi} with the
auxiliary field $\phi$ is equivalent to the original action \eqref{R2}.
Now, let us perform a Weyl rescaling of the metric
\begin{equation}
g_{\mu\nu}=e^{-\phi} \overline{g}_{\mu\nu}.
\end{equation}
It follows that
\begin{equation}
\begin{split}
\sqrt{-g} & = e^{-2\phi}\sqrt{-\overline{g}}\, , \\
\mathcal{R} & = e^{\phi}\left(\overline{\mathcal{R}} +
3\overline{\nabla}^{2} \phi - \frac{3}{2}
\left[\overline{\nabla}\phi\right]^{2}\right),
\end{split}
\end{equation}
where $\overline{\nabla}_{\lambda}\overline{g}_{\mu\nu}=0$. Therefore,
\begin{equation}
\begin{split}
\sqrt{-g}e^{\phi}\mathcal{R} &= \sqrt{- \overline{g}} \left(
\overline{R}+3\overline{\nabla}^2 \phi - \frac{3}{2}
\left[\overline{\nabla}\phi\right]^2\right), \\
-\frac{3}{2}m^{2}\sqrt{-g}\left(e^{\phi}-1\right)^{2} &=
-\frac{3}{2}m^{2} \sqrt{-\overline{g}} \left(1-e^{-\phi}\right)^{2},
\end{split}
\end{equation}
and the action becomes
\begin{equation}
\mathcal{S} = \int{d^{4}x \sqrt{-
\overline{g}}\left(\overline{\mathcal{R}}-\frac{3}{2}
\left[\overline{\nabla}\phi\right]^2-\frac{3}{2}m^{2}\left[1-e^{-\phi}
\right]^2 \right)},
\end{equation}
where we have dropped a total divergence term. It follows that the
higher-derivative pure-gravity theory described by action \eqref{R2} is
equivalent to a theory of normal Einstein gravity coupled to a real
scalar field $\phi$. It is important to note that, with respect to the
metric signature $(-,+,+,+)$ we are using, the kinetic energy term for
$\phi$ has the correct sign and, hence, that $\phi$ is not ghost like.
Also, note that a unique potential-energy function
\begin{equation}
V(\phi)=\frac{3}{2}m^2 \left(1-e^{-\phi}\right)^2
\end{equation}
emerges which has a stable minimum at $\phi=0$. We conclude that
$\mathcal{R}+\mathcal{R}^2$ gravitation with metric $g_{\mu\nu}$ is
equivalent to $\overline{\mathcal{R}}$ gravitation with metric
$\overline{g}_{\mu\nu}$ plus a non-ghost real scalar field $\phi$ with a
fixed potential energy and a stable vacuum state. The property that
$\phi$ is not ghost-like is sufficiently important that we will present
yet another proof of this fact. This proof was first presented in
\cite{B}. If we expand the metric tensor as
\begin{equation}
g_{\mu\nu}=\eta_{\mu\nu}+h_{\mu\nu},
\end{equation}
then the part of action \eqref{R2} quadratic in $h_{\mu\nu}$ is given by
\begin{equation}
\mathcal{S}=\int{d^{4}x\left[\frac{1}{4}h^{\mu\nu}\left(\nabla^{2}
\left\{ P_{\mu\nu\rho\sigma}^{(2)} - 2P_{\mu\nu\rho\sigma}^{(0)}
\right\} + 2m^{-2}\left(\nabla^{2} \right)^2
P_{\mu\nu\rho\sigma}^{(0)}\right)h^{\rho\sigma}\right]},
\end{equation}
where $P_{\mu\nu\rho\sigma}^{(2)}$ and $P_{\mu\nu\rho\sigma}^{(0)}$ are
transverse projection operators for $h_{\mu\nu}$. Inverting the kernel
yields the propagator
\begin{align}
\Delta_{\mu\nu\rho\sigma}^{-1} &=
\left(\nabla^{2}P_{\mu\nu\rho\sigma}^{(2)}
+ 2m^{-2}\nabla^{2}\left[ \nabla^{2}-m^{2}\right]P_{\mu\nu\rho\sigma}
\right)^{-1}
\nonumber \\
&= \frac{1}{\nabla^{2}}\left(P_{\mu\nu\rho\sigma}^{(2)}-
\frac{1}{2}P_{\mu\nu\rho\sigma}^{(0)}\right)+\frac{1}{2(\nabla^{2}-
m^{2})}P_{\mu\nu\rho\sigma}^{(0)}.
\end{align}
The term proportional to $(\nabla^2)^{-1}$ corresponds to the usual
two-helicity massless graviton. However, the term proportional to
$(\nabla^2-m^2)^{-1}$ represents the propagation of a real scalar field
with positive energy and, hence, not a ghost. This corresponds to the
results obtained using the auxiliary field above. We would like to point
out that there may be very interesting physics associated with the
scalar field $\phi$. For example, as emphasized in \cite{s1}, $\phi$ may
act as a natural inflaton in cosmology of the early universe.

Now let us consider Einstein gravity modified by quadratic terms
involving the Weyl tensor only. That is, consider the action
\begin{equation}
\mathcal{S} = \int{d^{4}x\sqrt{-g}\left(\mathcal{R}-\frac{1}{2}
m^{-2} C_{\mu\nu\alpha\beta} C^{\mu\nu\alpha\beta}\right)}.
\end{equation}
Using the identity
\begin{equation}
C_{\mu\nu\alpha\beta}C^{\mu\nu\alpha\beta}= \GB +
2\left(\mathcal{R}_{\mu\nu}\mathcal{R}^{\mu\nu}-
\frac{1}{3}\mathcal{R}^2\right),
\end{equation}
where $\GB$ is the topological Gauss-Bonnet combination defined in
\eqref{GB}, the action becomes
\begin{equation}
\mathcal{S}=\int{d^{4}x\sqrt{-g}\left(\mathcal{R}
-m^{-2}\left[\mathcal{R}_{\mu\nu} \mathcal{R}^{\mu\nu}-
\frac{1}{3}\mathcal{R}^2\right]\right)},
\label{Rmunu2}
\end{equation}
where we have dropped a total divergence. The fourth-order equations of
motion can be reduced to second order equations by introducing an
auxiliary symmetric tensor field $\phi_{\mu\nu}$. Using this field, the
action can be written as
\begin{equation}
{\mathcal{S}}=\int{d^{4}x\sqrt{-g}\left({\mathcal{R}}-
G_{\mu\nu}\phi^{\mu\nu} + \frac{m^{2}}{4}
\left[\phi_{\mu\nu}\phi^{\mu\nu}-\phi^{2}\right]\right)},
\label{Rphimunu}
\end{equation}
where $\phi=\phi_{\mu\nu}g^{\mu\nu}$ and
$G_{\mu\nu}=\mathcal{R}_{\mu\nu}-\frac{1}{2}g_{\mu\nu}\mathcal{R}$ is
the Einstein tensor. Note that the $\phi_{\mu\nu}$ equation of motion is
\begin{equation}
\phi_{\mu\nu}=2m^{-2}\left(\mathcal{R}_{\mu\nu}-
\frac{1}{6}g_{\mu\nu}\mathcal{R}\right).
\end{equation}
Substituting this into \eqref{Rphimunu} gives back the original action
\eqref{Rmunu2}. As it stands, action \eqref{Rphimunu} is somewhat
obscure since the $G_{\mu\nu}\phi^{\mu\nu}$ term mixes $g_{\mu\nu}$ and
$\phi_{\mu\nu}$ at the quadratic level. They can, however, be decoupled
by a field redefinition. First write the above action as
\begin{equation}
\mathcal{S}=\int{d^{4}x \sqrt{-g}
\left(\left[\left\{1+\frac{1}{2}\phi\right\}g^{\mu\nu}
-\phi^{\mu\nu} \right]
\mathcal{R}_{\mu\nu}+\frac{m^{2}}{4}\left[\phi_{\mu\nu}
\phi^{\mu\nu}-\phi^{2}\right]\right)}.
\end{equation}
Now transform the metric as
\begin{equation}
\sqrt{-\overline{g}}\overline{g}^{\mu\nu}=\sqrt{-g}
\left(\left[1+\frac{1}{2}\phi \right]g^{\mu\nu} -
{\phi^{\mu}}_{\alpha}g^{\alpha\nu}\right),
\end{equation}
or, equivalently,
\begin{equation}
\begin{split}
g_{\mu\nu} &= \left(\det A\right)^{-1/2} {A_\mu}^\alpha
\overline{g}_{\alpha\nu}, \\
{A_\mu}^\alpha &= \left(1+\frac{1}{2}\phi\right){\delta_\mu}^{\alpha}
-{\phi_\mu}^{\alpha}.
\end{split}
\end{equation}
Under this transformation
\begin{equation}
\mathcal{R}_{\mu\nu}=\overline{\mathcal{R}}_{\mu\nu} -
\overline{\nabla}_{\mu} {C^{\alpha}}_{\alpha\nu} +
\overline{\nabla}_{\alpha} {C^{\alpha}}_{\mu\nu}+{C^{\alpha}}_{\mu\nu}
{C^{\beta}}_{\alpha\beta} -
{C^{\alpha}}_{\mu\beta}{C^{\beta}}_{\nu\alpha},
\end{equation}
where $\overline{\nabla}_{\lambda}\overline{g}_{\mu\nu}=0$ and
\begin{equation}
\begin{split}
{C^{\alpha}}_{\mu\nu} &= \frac{1}{2}\left(X^{-
1}\right)^{\alpha\beta}\left(\overline{\nabla}_{\mu}
X_{\nu\beta}+\overline{\nabla}_{\nu}X_{\mu\beta} -
\overline{\nabla}_\beta X_{\mu\nu}\right), \\
X_{\mu\nu} &= g_{\mu\nu}=\left(\det A\right)^{-1/2}{A_\mu}^\alpha
\overline{g}_{\alpha\nu}.
\end{split}
\end{equation}
Inserting these transformations into the above and dropping a total
divergence, the action becomes \cite{s2}
\begin{equation}
\mathcal{S}=\int{d^{4}x\sqrt{-
\overline{g}}\left[\overline{\mathcal{R}}+\overline{g}^
{\mu\nu}\left({C^{\alpha}}_{\mu\nu}{C^{\beta}}_{\beta\alpha}-
{C^{\alpha}}_{\mu\beta}
{C^{\beta}}_{\nu\alpha}\right)-\frac{m^{2}}{4}\left(\det A\right)^{-
1/2}\left(\phi_{\mu\nu}\phi^{\mu\nu}-\phi^{2}\right)\right]}.
\label{Rbarphimunu}
\end{equation}
Note that the action for $\phi_{\mu\nu}$ is a complicated non-linear
sigma model since $C=C(X)$ and $X=X(\phi)$. It is useful to consider the
kinetic energy part of the action expanded to quadratic order in
$\phi_{\mu\nu}$ only. It is found to be
\begin{equation}
\mathcal{S}^{\text{quad}}_\phi = \int d^4 x \sqrt{-\overline{g}} \left(
\frac{1}{4} \overline{\nabla}^\alpha \phi^{\mu\nu}
\overline{\nabla}_\alpha \phi_{\mu\nu}
-\frac{1}{2} \overline{\nabla}^\alpha \phi^{\mu\nu}
\overline{\nabla}_\mu \phi_{\nu\alpha} + \frac{1}{2}
\overline{\nabla}_\mu \phi^{\mu\nu} \overline{\nabla}_\nu \phi
-\frac{1}{4} \overline{\nabla}^\alpha \phi \overline{\nabla}_\alpha \phi
\right).
\end{equation}
This action is clearly the curved space generalization of the
Pauli-Fierz action for a spin-2 field except that every term has the
wrong sign! This implies, of course, that $\phi_{\mu\nu}$ propagates as
a ghost. It is interesting to note that the kinetic energy and curvature
tensor in the action \eqref{Rbarphimunu} are invariant under the gauge
transformation
\begin{equation}
\begin{split}
\phi'_{\mu\nu} &= \phi_{\mu\nu}+\overline{\nabla}_{(\mu}\xi_{\nu)}
-C^{\alpha}_{\mu\nu}\xi_{\alpha}, \\
 \overline{g}'_{\mu\nu} &= \left(1+\frac{1}{2}\phi'\right)^{-1}
\left[\frac {\det^{1/2}A'}{\det^{1/2}A} \left(1+\frac{1}{2}\phi\right)
\overline{g}_{\mu\nu} + \left(\phi'_{\mu\nu}-\phi_{\mu\nu}\right)
\right]. \\
\end{split}
\end{equation}
This insures that the above action describes a consistent coupling of a
spin-2 symmetric tensor field $\phi_{\mu\nu}$ to Einstein gravitation at
the full non-linear level \cite{s3}. We conclude, therefore, that
$\mathcal{R}+C^{2}$ gravitation with metric $g_{\mu\nu}$ is equivalent
to $\overline{\mathcal{R}}$ gravity with metric $\overline{g}_{\mu\nu}$
plus a ghost-like symmetric tensor field $\phi_{\mu\nu}$ with a
consistent non-linear coupling to gravity and a fixed potential energy.
The physics in the field $\phi_{\mu\nu}$ is obscured by its ghost-like
nature. However, this can be altered by yet
higher-derivative terms, such as those one would expect to find
generated in superstring theories. Therefore, at long last, we turn to
our discussion of quadratic supergravitation in superstring theory.

\section{Superspace Formalism}

In the K{\"a}hler (Einstein frame) superspace formalism, the most
general Lagrangian for Einstein gravity, matter and gauge fields is
\begin{equation}
\mathcal{L}_E=-
\frac{3}{2\kappa^{2}}\int{d^{4}{\theta}E[K]}+\frac{1}{8}\int{d^{4}
\theta\frac{E}{R}f(\Phi_{i})_{ab}W^{{\alpha}a} {W_\alpha}^b} +
\text{h.c.},
\label{eq:fifth}
\end{equation}
where we have ignored the superpotential term which is irrelevant for
this discussion. The fundamental supergravity superfields are $R$ and
$W_{\alpha\beta\gamma}$, which are chiral, and $G_{\alpha{\dot\alpha}}$,
which is Hermitian. The bosonic $\mathcal{R}^2$, $(C_{mnpq})^{2}$ and
$(\mathcal{R}_{mn})^{2}$ terms are contained in the highest components
of the superfields $\bar{R}R$, $(W_{\alpha\beta\gamma})^{2}$ and
$(G_{\alpha\dot{\alpha}})^{2}$ respectively. One can also define the
superGauss-Bonnet combination
\begin{equation}
\SGB = 8(W_{\alpha\beta\gamma})^{2}+(\bar{\mathcal{D}}^2-
8R)(G_{\alpha\dot{\alpha}}^2-4\bar{R}R).
\end{equation}
The bosonic Gauss-Bonnet term is contained in the highest chiral
component of $\SGB$. It follows that the most general quadratic
supergravity Lagrangian is given by
\begin{equation}
\mathcal{L}_Q = \int d^4 \theta E \left[
\Sigma(\bar{\Phi}_i,\Phi_i) \bar{R} R +
\frac{1}{R} g(\Phi_i) (W_{\alpha\beta\gamma})^2
+ \Delta(\bar{\Phi}_i,\Phi_i) (G_{\alpha\dot{\alpha}})^2 + \text{h.c.}
\right].
\end{equation}

Although our discussion is perfectly general, we will limit ourselves to
orbifolds, such as $Z_{4}$, which have $(1,1)$ moduli only. The relevant
superfields are the dilaton, $S$, the diagonal moduli $T^{II}$, which we
will denote as $T^{I}$, and all other moduli and matter superfields,
which we denote collectively as $\phi^{i}$. The associated K{\"a}hler
potential is
\begin{equation}
\begin{split}
K &= K_{0} + Z_{ij}\bar{\phi}^{i}\phi^{j} +
{\mathcal{O}}((\bar{\phi}\phi)^{2}), \\
\kappa^{2}K_{0} &= -\ln(S+\bar{S})-\sum{(T^{I}+\bar{T}^{I})}, \\
Z_{ij} &= \delta_{ij} \prod{(T^{I}+\bar{T}^{I})^{q^{i}_I}}.
\end{split}
\end{equation}
The tree level coupling functions $f_{ab}$ and $g$ can be computed
uniquely from amplitude computations and are given by
\begin{equation}
f_{ab} = \delta_{ab}k_{a}S, \qquad g = S.
\end{equation}
There is some ambiguity in the values of $\Delta$ and $\Sigma$ due to
the ambiguity in the definition of the linear supermultiplet. We will
take the conventional choice
\begin{equation}
\Delta=-S, \qquad \Sigma=4S.
\end{equation}
It follows that, at tree level, the complete $Z_{N}$ orbifold Lagrangian
is given by $\mathcal{L}=\mathcal{L}_E+\mathcal{L}_Q$ where
\begin{equation}
\mathcal{L}_Q = \frac{1}{4}\int{d^4\theta\frac{E}{R}S\, \SGB}+
\text{h.c.}
\end{equation}
Using this Lagrangian, we now compute the one-loop
moduli-gravity-gravity anomalous threshold correction \cite{E}. This
must actually be carried out in the conventional (string frame)
superspace formalism and then transformed to K{\"a}hler superspace
\cite{F}. We also compute the relevant superGreen-Schwarz graphs. Here
we will simply present the result. We find that
\begin{align}
\mathcal{L}_{\mathrm{massless}}^{\text{1-loop}} = &
\frac{1}{24(4\pi)^{2}} \sum \left[h^{I}\int{d^{4}\theta
(\bar{\mathcal{D}}^{2} - 8
R)\bar{R}R\frac{1}{\partial^{2}}D^{2}\ln(T^{I}+\bar{T}^{I})} \right.
\nonumber \\
&+ (b^{I}-8p^{I}) \int{d^{4}\theta(W_{\alpha\beta\gamma})^{2}
\frac{1}{\partial^{2}} D^{2}\ln(T^{I}+\bar{T}^{I})}
\nonumber \\
&\left. +p^{I}\int{d^{4}\theta(8(W_{\alpha\beta\gamma})^{2}+
(\bar{\mathcal{D}}^2 - 8R)((G_{\alpha\dot{\alpha}})^{2}-
4\bar{R}R))\frac{1}{\partial^{2}}
D^{2}\ln(T^{I}+\bar{T}^{I})}+ \text{h.c.} \right],
\end{align}
where
\begin{equation}
\begin{split}
h^I &= \frac{1}{12}(3\gamma {T}+3\vartheta_{T}q^{I}+\varphi), \\
b^I &= 21+1+n_{M}^{I}-\dim G + \sum (1+2q_{I}^{i})-24\delta_{GS}^{I}, \\
p^{I} &= -\frac{3}{8}\dim G-\frac{1}{8}-\frac{1}{24}\sum 1+\xi-3
\delta_{GS}^{I}.
\end{split}
\end{equation}
The coefficients $\gamma_{T}$ and $\vartheta_{T}$, which arise from
moduli loops, and $\varphi$ and $\xi$, which arise from gravity and
dilaton loops, are unknown. However, as we shall see, it is not
necessary to know their values to accomplish our goal. Now note that if
$h^{I}\neq0$ then there are non-vanishing $\mathcal{R}^2$ terms in the
superstring Lagrangian. If $b^{I}-8p^{I}\neq0$ then the Lagrangian has
$C^{2}$ terms. Coefficient $p^{I}\neq0$ merely produces a Gauss-Bonnet
term. With four unknown parameters what can we learn? The answer is, a
great deal! Let us take the specific example of the $Z_{4}$ orbifold. In
this case, the Green-Schwarz coefficients are known \cite{G}
\begin{equation}
\delta_{GS}^{1,2}=-30, \qquad \delta_{GS}^{3}=0,
\end{equation}
which gives the result
\begin{equation}
b^{1,2}=0, \qquad b^{3}=11\times24.
\end{equation}
Now, let us try to set the coefficients of the
$(C_{\mu\nu\alpha\beta})^{2}$ terms to zero simultaneously. This implies
that
\begin{equation}
b^{I}=8p^{I}
\end{equation}
for $I=1,2,3$ and therefore that
\begin{equation}
p^{1,2}=0, \qquad p^{3}=33.
\end{equation}
{}From this one obtains two separate equations for the parameter $\xi$
given by
\begin{equation}
\xi=\frac{3}{8}\dim G+\frac{1}{8}+\frac{1}{24}\sum 1-90,
\end{equation}
for $I=1,2$ and
\begin{equation}
\xi=\frac{3}{8}\dim G+\frac{1}{8}+\frac{1}{24}\sum 1,
\end{equation}
for $I=3$. Clearly these two equations are incompatible and, hence, it
is impossible to have all vanishing $(C_{\mu\nu\alpha\beta})^{2}$ terms
in the 1-loop corrected Lagrangian of $Z_{4}$ orbifolds. We find that
the same results hold in other orbifolds as well.

\section{Supersymmetry Breaking in $D=2$ Supergravity}

Having demonstrated that quadratic gravitational terms can appear in
four-dimensional, $N=1$ superstring Lagrangians, we would like to
consider the effect of such terms on the vacuum state. Although our
ultimate goal is to do this in superstring theory, at the present time
we content ourselves with exploring the same issue in quadratic $N=1$
supergravity theories, which are simpler and less constrained. Here, we
will present our results for $D=2$, $(1,1)$ supergravity. However, we
have shown that similar results generically occur in $D=4$, $N=1$
supergravity theories as well. The two-dimensional, $(1,1)$ supergravity
multiplet is composed of a graviton $g_{mn}$, a gravitino
${\chi_{m}}^\alpha$ and a real auxiliary scalar field $A$ \cite{N1}. The
relevant superfields are the superdeterminant $E$ given by
\begin{equation}
E=e\left( 1 + \frac{i}{2} \theta^{\alpha} \udu\gamma{m}\alpha\beta
\chi_{m\beta} + \overline{\theta}\theta \left[\frac{i}{4}A + \frac{1}{8}
\widetilde{\epsilon}^{mn} {\chi_m}^\alpha \udu\gamma5\alpha\beta
\chi_{n\beta}\right]\right)
\end{equation}
and a scalar superfield $S$, where
\begin{equation}
S=A + \theta^{\alpha}\psi_{\alpha}+\frac{i}{2} \overline{\theta}\theta C
\end{equation}
and
\begin{equation}
\begin{split}
C &= -{\mathcal{R}}-\frac{1}{2} {\chi_m}^\alpha \udu\gamma{m}\alpha\beta
\psi_{\beta}
+ \frac{i}{4} \widetilde{\epsilon}^{mn} {\chi_m}^{\alpha}
\udu\gamma5\alpha\beta \chi_{n\beta} A-\frac{1}{2}A^{2}, \\
\psi_{\alpha} &= -2i \widetilde{\epsilon}^{mn} \udu\gamma5\alpha\beta
{\mathcal{D}}_{m}\chi_{n \beta} - \frac{i}{2} \udu\gamma{m}\alpha\beta
\chi_{m\beta}A.
\end{split}
\end{equation}
The usual Einstein supergravity is described by the action
\begin{equation}
{\mathcal{S}}_{E}=2i\int{d^{2}xd^{2}\theta ES}.
\end{equation}
In component fields this simply becomes
\begin{equation}
{\mathcal{S}}_{E}=\int{d^{2}x e{\mathcal{R}}},
\end{equation}
which is a total divergence. That is, Einstein supergravity in two
dimensions has no propagating degrees of freedom and is purely
topological. The most general quadratic supergravity action is given by
\begin{equation}
{\mathcal{S}}_{E+Q}=2i\int{d^{2}xd^{2}\theta
E\left(f(S)+g(S){\mathcal{D}}^{\alpha}S
{\mathcal{D}}_{\alpha}S\right)},
\label{fg}
\end{equation}
where $f$ and $g$ are arbitrary functions of superfield $S$. Recall that
in two dimensions the Weyl tensor vanishes, so this theory contains
powers of the bosonic scalar curvature ${\mathcal{R}}$ only.
Furthermore, the structure of action \eqref{fg} is such that all higher
powers ${\mathcal{R}}^{n}$ for $n\geq3$ vanish, and there are never more
than two derivatives acting on the component field $A$. Here, for
simplicity, we will consider the special case where
\begin{equation}
\begin{split}
f(S)&=a+S+bS^{2}+dS^{3}, \\
g(S)&=ic,
\end{split}
\end{equation}
and $a,b,c$, and $d$ are real constants. In analogy with the bosonic
case discussed in Section 1, we introduce two auxiliary scalar
superfields, $\Lambda$ and $\Phi$. The above Lagrangian is then
equivalent to
\begin{equation}
{\mathcal{L}} = 2iE \left[a+S+b\Lambda^{2}+ic{\mathcal{D}}^{\alpha}
\Lambda{\mathcal{D}}_{\alpha} \Lambda+d\Lambda^{3} + \left(e^{\Phi}-
1\right)\left(S-\Lambda\right)\right].
\end{equation}
Inserting the equations of motion for $\Lambda$ and $\Phi$ back into
this Lagrangian yields the original action \eqref{fg}. Again, in analogy
with the bosonic case, we perform a superWeyl transformation of the form
\begin{equation}
\begin{split}
\widetilde{E} &= e^{\Phi}E, \\
\widetilde{S} &= e^{-\Phi}S+ie^{-
\Phi}{\mathcal{D}}^{\alpha}{\mathcal{D}}_{\alpha}\Phi.
\end{split}
\end{equation}
The Lagrangian then takes the form
\begin{equation}
{\mathcal{L}}=2iE\left[e^{\Phi}S+ie^{\Phi}{\mathcal{D}}^{\alpha}
\Phi{\mathcal{D}}_{\alpha} \Phi+e^{-\Phi}\left(1-
e^{\Phi}\right)\Lambda+ae^{-\Phi}+be^{-\Phi}\Lambda^{2} +
ic{\mathcal{D}}^{\alpha}\Lambda{\mathcal{D}}_{\alpha}\Lambda + de^{-
\Phi}\Lambda^{3}\right],
\label{SPL}
\end{equation}
where we have dropped the tilde. It follows that quadratic supergravity
is equivalent to Einstein supergravity (modified by the $e^{\Phi}$
factor in front of $S$) coupled to two new scalar superfield degrees of
freedom. Superfields $\Lambda$ and $\Phi$ can be expanded into component
fields as
\begin{equation}
\begin{split}
\Lambda &= \lambda+i\theta^{\alpha}\zeta_{\alpha}+\frac{i}{2}
\overline{\theta}\theta G, \\
\Phi &= \phi+i\theta^{\alpha}\pi_{\alpha}+\frac{i}{2}
\overline{\theta}\theta F.
\end{split}
\end{equation}
Inserting these expressions, and the expansion of S, into Lagrangian
\eqref{SPL} and eliminating the auxiliary fields $A$, $G$ and $F$ yields
a component field Lagrangian of the form
\begin{equation}
\mathcal{L} = \mathcal{L}_{\text{KE-Boson}} - e V(\phi,\lambda)
+ \mathcal{L}_{\text{KE-Fermion}} + \mathcal{L}_{\text{M-Fermion}}
+ \mathcal{L}_{\text{Boson-Fermion}}.
\end{equation}
The boson kinetic energy term is given by
\begin{equation}
\mathcal{L}_{\text{KE-Boson}}= e \left\{ e^{\phi}{\mathcal{R}}-
2e^{\phi}\left(\nabla_{m}\phi\right)^{2} -2
c\left(\nabla_{m}\lambda\right)^{2} \right\}.
\end{equation}
Note that both $\lambda$ and $\phi$ are physically propagating scalar
fields. The potential energy term is found to be
\begin{align}
V &= \frac{1}{8c}\left( 1-2e^{-\phi} \left[ 1+2b \lambda
+\left(2c+3d\right)\lambda^{2}\right] + e^{-2\phi} \left[1 +
4\left(b+ac\right) \lambda + 2\left(2b^{2} + 2c+3d\right)\lambda^{2}
\right.\right. \notag \\
& \hspace*{0.15in} \left.\left. + 4b \left(c+3d\right)
\lambda^{3}+d\left(4c+9d\right) \lambda^{4}\right]\right).
\end{align}
We now solve generically for extrema of this potential with vanishing
cosmological constant. We find that all such extrema,
$(\lambda_{0},\phi_{0})$, satisfy
\begin{equation}
\begin{split}
a &= \frac{4}{27}\frac{b^{3}}{\left(c+2d\right)^{2}}, \\
\lambda_0 &= \frac{-2b}{3\left(c+2d\right)}, \\
1-e^{-\phi_0} &= \frac{-4b^{2} \left(c+3d\right) }{9c^{2}-
4b^{2}\left(c+3d\right) +36d\left(c+d\right)}.
\end{split}
\end{equation}
Evaluated at these extrema, the fermion mass term in the Lagrangian is
given by
\begin{align}
\mathcal{L}_{\text{M-Fermion}} & =
m_{11}i\pi^{\alpha}\pi_{\alpha}+m_{22} i \zeta^{\alpha}
\zeta_{\alpha}+m_{33} \widetilde{\epsilon}^{mn} {\chi_m}^{\alpha}
\udu\gamma5\alpha\beta \chi_{n\beta} +
2m_{12}i\pi^{\alpha}\zeta_{\alpha} \notag \\
&\hspace*{0.2in} + 2m_{13}i\pi^{\alpha}\udu\gamma{m}\alpha\beta
\chi_{m\beta} + 2m_{23}i\zeta^{\alpha}\udu\gamma{m}\alpha\beta
\chi_{m\beta},
\end{align}
where
\begin{equation}
\begin{split}
m_{11} &= \frac{-2b\left(-2b^{2}+3c+6d\right)}{P}, \\
m_{22} &= \frac{9bc\left(c+2d\right)}{P}, \\
m_{33} &= \frac{2b^{3}c}{3\left(c+2d\right)P}, \\
m_{12} &= \frac{-3c\left(3c-4b^{2}+12d\right)+12d\left(b^{2}-
3d\right)}{2P}, \\
m_{13} &= \frac{bc\left(9c-8b^{2}+36d\right)-12bd\left(b^{2}-3d\right)}
{6\left(c+2d\right)P}, \\
m_{23} &= \frac{-2b^{2}c}{P},
\label{masses}
\end{split}
\end{equation}
where $P$ is a polynomial in $b,c$, and $d$ given by
\begin{equation}
P=c\left(9c-4b^{2}+36d\right)-12d\left(b^{2}-3d\right).
\end{equation}
Some ranges of parameters $b,c,d$ correspond to $(\lambda_{0},\phi_{0})$
being a local maximum or a saddle point. However, for a large range of
parameters, we find that $(\lambda_{0},\phi_{0})$ is a local minimum.
Furthermore, this minimum is very stable against quantum tunneling since
the potential energy barrier around it is of the order of the Planck
scale. Let us evaluate the fermion mass matrix for $b,c$, and $d$
corresponding to such a minimum, and then diagonalize it. The result is
that, in a new fermion basis labeled by $\widetilde{\chi}_{m}^{\alpha},
\widetilde{\pi}$ and $\widetilde{\zeta}$, the square of the fermion mass
matrix is
\begin{equation}
\widetilde{M}^{\dagger}\widetilde{M}=\left(\begin{array}{rcl}
m_{33}^{2}& & \\
          &0& \\
          & &\widetilde{m}^{2}
\end{array}\right),
\end{equation}
where $m_{33}$ and $\widetilde{m}$ are non-vanishing, and $m_{33}$ is
given in equation \eqref{masses}. Note the vanishing mass for
$\widetilde{\pi}$. This implies that $\widetilde{\pi}$ is a Goldstone
fermion and, hence, that supersymmetry is spontaneously broken at this
vacuum. This conclusion is further strengthened by the fact that the
gravitino, $\widetilde{\chi}_{m}^{\alpha}$, has acquired a non-vanishing
mass. As a final check that supersymmetry is indeed broken, one can
compute the supersymmetry transformation of the three diagonal fermions,
evaluated at the vacuum. Schematically, we find that
\begin{equation}
\begin{split}
\delta_{SUSY}\widetilde{\chi}_{m}^{\alpha} &= \cdots + 0, \\
\delta_{SUSY}\widetilde{\pi} &= \cdots + \text{non-zero}, \\
\delta_{SUSY}\widetilde{\zeta} &= \cdots + 0.
\end{split}
\end{equation}
The inhomogeneous term in the supersymmetry transformation of
$\widetilde{\pi}$, proves that this vacuum spontaneously breaks
supersymmetry and, in fact, is the reason why $\widetilde{\pi}$ is a
massless Goldstone fermion.

We conclude that in two-dimensional $(1,1)$ quadratic supergravity there
exist, for a large range of parameters, stable vacua with vanishing
cosmological constant that spontaneously break the $(1,1)$
supersymmetry. Supersymmetry is broken by two new scalar superfield
degrees of freedom that are contained in the supervielbein in quadratic
supergravity. Importantly, this result is not restricted to two
dimensions. We now show that exactly the same phenomenon occurs in
$D=4$, $N=1$ quadratic supergravitation.

\section{Supersymmetry Breaking in Four Dimensions}

The Einstein supergravity action in the conventional $N=1$ superspace
formalism is given by
\begin{equation}
{\mathcal{S}}_{E}=-3\int{d^{4}xd^{4}\theta E}.
\end{equation}
We now want to generalize this result to include higher derivative
gravitational terms. We find that the most general higher derivative
supergravity theory involving ${\mathcal{R}}^{2}$ is described by the
action
\begin{equation}
{\mathcal{S}}_{E+R^{2}}=\int{d^{4}xd^{4}\theta E f\left(R^{\dagger},
R\right)},
\label{fRR}
\end{equation}
where $f$ is an arbitrary real function. This expression can be broken
into the sum of a left chiral and a right chiral integral as follows.
\begin{equation}
{\mathcal{L}}=-\frac{1}{8}\int{d^{2}\Theta
2{\mathcal{E}}\left(\bar{\mathcal{D}}^{2}-8R\right)f}-
\frac{1}{8}\int{d^{2}\bar\Theta
2\bar{\mathcal{E}}\left({\mathcal{D}}^{2}-8R^{\dagger}\right)f}.
\end{equation}
Suppressing all fermion component fields, the superfields
${\mathcal{E}}$ and $R$ are given by
\begin{equation}
\begin{split}
2{\mathcal{E}} &= e\left(1-\Theta^{2}M^{\dagger}\right), \\
R &= -\frac{1}{6}\left(M+\Theta^{2}\left[-
\frac{1}{2}{\mathcal{R}}+\frac{2}{3}M^
{\dagger}M+\frac{1}{3}b_{a}b^{a}-i{\mathcal{D}}_{a}b^{a}\right]\right),
\end{split}
\end{equation}
where the graviton is contained in $\mathcal{R}$, and $M$ is the complex
scalar field and $b_{a}$ the real vector field of the gravity
supermultiplet. Expanding the above Lagrangian in component fields,
again suppressing all fermionic fields, we find that
\begin{align}
{\mathcal{L}} = & \frac{e}{6} \left\{ \left(f-
\frac{1}{18}f_{AA^{\dagger}}\left[2M^{\dagger}M+
b_{a}b^{a}\right]-
\frac{1}{6}\left[f_{A}M+f_{A^{\dagger}}M^{\dagger}\right]\right)
{\mathcal{R}}+\frac{1}{24}f_{AA^{\dagger}}{\mathcal{R}}^{2} \right.
\notag \\
& \left. \qquad -\frac{1}{6}f_{AA^{\dagger}} \partial_{m}
M^{\dagger}\partial^{m}M+\frac{1}{6}f_{AA^{\dagger}}{\mathcal{D}}_{a}
b^{a}{\mathcal{D}}_{c}b^{c}+\cdots \right\},
\end{align}
where $A=-\frac{1}{6}M$. Note that since the Lagrangian contains
${\mathcal{R}}+{\mathcal{R}}^{2}$ bosonic gravity, there is an extra
real scalar degree of freedom in the metric tensor that now propagates.
Also, we see that the complex field $M$ and one component of the real
vector field $b_{a}$, which for Einstein supergravity are auxiliary
fields, now begin to propagate. It follows that the above class of
higher-derivative supergravity theories have, in addition to the usual
helicity-two graviton, four new real degrees of freedom. These could
only arrange themselves into two chiral multiplets. Therefore, we are
led to introduce two Lagrange multiplier chiral superfields $\Lambda$
and $\Phi$. In terms of these, Lagrangian \eqref{fRR} can be written as
\begin{equation}
{\mathcal{L}} = \int{d^{4}\theta
E\left(f\left(\Phi^{\dagger},\Phi\right)+\Lambda+\Lambda^{\dagger}-
\frac{\Lambda \Phi}{R}-
\frac{\Lambda^{\dagger}\Phi^{\dagger}}{R^{\dagger}}\right)}.
\label{PhiLam}
\end{equation}
To check this, let us deduce the superfield equation of motion of
$\Lambda$. This is found from
\begin{align}
\delta_{\Lambda}{\mathcal{L}} &= \int{d^{4}\theta E\left(\delta\Lambda-
\delta\Lambda \frac{\Phi}{R}\right)} \notag \\
&= -\frac{1}{4}\int{d^{2}\Theta
2{\mathcal{E}}\left(\bar{{\mathcal{D}}}^{2}-8R\right)\left
(\delta\Lambda-\delta\Lambda\frac{\Phi}{R}\right)} \notag \\
&= -\frac{1}{4}\int{d^{2}\Theta 2{\mathcal{E}}\left(-
8R+8\Phi\right)\delta\Lambda}.
\end{align}
Setting $\delta_{\Lambda}{\mathcal{L}}$ to zero yields the equation of
motion
\begin{equation}
\Phi=R.
\end{equation}
Substituting this back into \eqref{PhiLam}, one finds the original
action \eqref{fRR}. For completeness, we display the $\Phi$ equation of
motion which is given by
\begin{equation}
\Lambda=-\frac{1}{8}\left(\bar{\mathcal{D}}^{2}-8R\right)\frac{\partial
f}{\partial \Phi}.
\end{equation}
Now let us compare Lagrangian \eqref{PhiLam} with the standard form of
the Lagrangian for chiral matter coupled to supergravity in conventional
superspace. This is given by
\begin{equation}
{\mathcal{L}}=\int{d^{4}\theta
E\left(-3e^{-
\frac{K}{3}}+\frac{W}{2R}+\frac{W^{\dagger}}{2R^{\dagger}}\right)},
\end{equation}
where $K$ and $W$ are the K\"{a}hler potential and the superpotential
respectively. It follows that $\Lambda$ and $\Phi$ couple to Einstein
supergravity with the specific K\"{a}hler and superpotentials
\begin{equation}
\begin{split}
K &= -3\ln\left[-
\frac{1}{3}\left(f\left(\Phi^{\dagger},\Phi\right)+\Lambda+
\Lambda^{\dagger}\right)\right], \\
W &= 6 \Phi\Lambda.
\end{split}
\end{equation}
Therefore, we conclude that general ${\mathcal{R}}^{2}$ supergravitation
is equivalent to two chiral supermultiplets coupled to Einstein
supergravity with specific K\"{a}hler and superpotentials. Let us now,
ignoring fermionic component fields, expand out the $\Lambda$ and $\Phi$
superfields. These are given by
\begin{equation}
\begin{split}
\Lambda&=B+\Theta^{2}G, \\
\Phi&=A+\Theta^{2}F.
\end{split}
\end{equation}
Inserting these into the above Lagrangian, eliminating all the auxiliary
fields and Weyl rescaling yields the bosonic Lagrangian
\begin{equation}
\frac{{\mathcal{L}}}{e}=-\frac{1}{2}{\mathcal{R}}-
K_{A^{\dagger}A}\partial_{m}
A^{\dagger}\partial^{m}A-
K_{B^{\dagger}B}\partial_{m}B^{\dagger}\partial^{m}B
-V\left(A,B\right),
\end{equation}
where the potential energy $V$ is given by
\begin{align}
V & =e^{K}\left(K^{i\bar{j}}\left(D_{i}W\right)\left(D_{\bar{j}}W\right)
-3\left|W\right|^{2}\right) \notag \\
&= \frac{4}{9}
\left(f\left(A^{\dagger},A\right)+B+B^{\dagger}\right)\left[
f^{-1}_{A^{\dagger}A}\left|B-
f_{A}A+2f_{A^{\dagger}A}A^{\dagger}A\right|^{2} \right. \notag \\
& \left. \hspace*{2in} -\left|A\right|^{2}\left(f-
2\left(f_{A}A+f_{A^{\dagger}}A^{\dagger}\right)+4
f_{A^{\dagger}A}A^{\dagger}A\right)\right].
\end{align}
Can we find a stable minimum with zero cosmological constant and
physical propagation for the $A$ and $B$ scalar fields that
spontaneously breaks supersymmetry? The answer is yes, as we will now
demonstrate. Consider the following example. Take
\begin{equation}
f\left(R^{\dagger},R\right)=-3\left(1-2\lambda^{-2}R^{\dagger}R+
\frac{\lambda^{-4}}{9}\left(R^{\dagger}R\right)^{2}\right).
\end{equation}
Then the associated k\"{a}hler and superpotentials are
\begin{equation}
\begin{split}
K &= -3 \ln\left(1-2\lambda^{-2}\Phi^{\dagger}\Phi+\frac{\lambda^{-
4}}{9}\Phi^{\dagger2} \Phi^{2}-\frac{1}{3}\Lambda-
\frac{1}{3}\Lambda^{\dagger}\right), \\
W &= 6 \Phi\Lambda.
\end{split}
\end{equation}
In component fields the potential energy becomes
\begin{multline}
V =  \frac{2}{27}\left(1-2\lambda^{-
2}\left|A\right|^{2}+\frac{1}{9}\lambda^{-4}
\left|A\right|^{4}-\frac{1}{3}B^{\dagger}-\frac{1}{3}B\right)^{-2}\left[
\lambda^{4}\left(9\lambda^{2}-2\left|A\right|^{2}\right)^{-1} \times
\right. \\
\left|B + \left. 2\lambda^{-
4}\left|A\right|^{2}\left(3\left|A\right|^{2}-
\lambda^{2}\right)
\right|^{2}+2\left|A\right|^{2}\left(\left|A\right|^{2}-
\lambda^{2}\right)^{2}
\right].
\end{multline}
This potential has stable minima with vanishing cosmological constant at
\begin{equation}
\begin{split}
\langle A \rangle &=0, \lambda e^{i\theta}, \\
\langle B \rangle &= -2\lambda^{-
4}\left|A\right|^{2}\left(3\left|A\right|^{2}-
\lambda^{2}\right).
\end{split}
\end{equation}
It is straightforward to show that both the $A$ and $B$ fields propagate
physically around these vacua. For the minimum at $\langle A
\rangle=\langle B \rangle=0$, the K\"{a}hler covariant derivatives of
$W$ with respect to $A$ and $B$ both vanish. It follows that
supersymmetry remains unbroken at this minimum. However, for the minimum
at $\langle A \rangle=\lambda e^{i\theta}$ and $\langle B \rangle=-4$,
the K\"{a}hler covariant derivatives are given by
\begin{equation}
\begin{split}
\langle D_{A}W \rangle &= -16, \\
\langle D_{B}W \rangle &= \frac{5}{2}\lambda e^{i\theta}.
\end{split}
\end{equation}
It follows that supersymmetry is indeed spontaneously broken at this
minimum. Note that the gravitino mass is
\begin{equation}
m_{\frac{3}{2}}=\langle e^{K} \left|W\right|^{2}
\rangle=\left(\frac{3}{2}\right)^{6}\lambda^{2}.
\end{equation}
We conclude that supersymmetry is generically broken in the vacua of
higher-derivative supergravitation theories. This mechanism is presently
being applied to realistic models of particle physics and will be
reported on elsewhere \cite{end}.

\end{document}